\documentstyle[aps,multicol,epsf]{revtex}
\draft
\begin{document}
        \title{Quantum noise in the position measurement
        of a cavity mirror undergoing Brownian motion}

\author{K. Jacobs\footnote{Present address: T-8, Theoretical Division, 
MS-B285,  Los Alaoms National Laboratory, Los Alamos, New Mexico 87545. Email: k.jacobs@lanl.gov}}
\address{Optics Section, The Blackett Laboratory, Imperial College,
London SW7
 2BZ, England}
\author{I. Tittonen\footnote{Present address: Helsinki University of
Technology, Metrology Research Institute, P. O. Box 3000, FIN-02015
HUT, Finland. Email: ilkka.tittonen@hut.fi}}
\address{Fakult\"{a}t Physik, Universit\"{a}t Konstanz, Konstanz
D-78457, Germany}
 \author{H. M. Wiseman}
\address{Department of Physics, The University of Queensland, St.
Lucia 4072, Australia}
\author{S. Schiller}
\address{Fakult\"{a}t Physik, Universit\"{a}t Konstanz, Konstanz
D-78457, Germany}
\maketitle

\begin{abstract}
We perform a quantum theoretical calculation of the noise power
spectrum for a phase measurement of the light output from a coherently
driven optical cavity with a freely moving rear mirror. We examine how
the noise resulting from the quantum back action appears among the
various contributions from other noise sources. We do not assume an
ideal (homodyne) phase measurement, but rather consider phase
modulation detection, which we show has a different shot noise level. 
We also take into account the effects of thermal damping of the mirror, 
losses within the cavity, and classical laser noise. We relate our
theoretical results to experimental parameters, so as to make direct
comparisons with current experiments simple. We also show that in this
situation, the standard Brownian motion master equation is inadequate
for describing the thermal damping of the mirror, as it produces a
spurious term in the steady-state phase fluctuation spectrum. The
corrected Brownian motion master equation [L. Diosi, Europhys.  Lett.
{\bf 22}, 1 (1993)] rectifies this inadequacy.
\end{abstract}

\pacs{42.50.lc,42.50.Dv,03.65.Bz,06.30.Bp}
\begin{multicols}{2}

\section{Introduction}
Interferometers provide a very sensitive method for detecting small
changes in the position of a mirror. This has been analysed extensively
in the context of gravitational wave
detection~\cite{gravwd,gravwd2,pace,gravwd3,gravwd4,gravwd5} and atomic
force microscopes~\cite{MJW,atmexp}. A key limit to the sensitivity of
such position detectors comes from the Heisenberg uncertainty principle.
The reduction in the uncertainty of the position resulting from the
measurement is accompanied by an increase in the uncertainty in momentum.
This uncertainty is then fed back into the position by the dynamics of
the object being measured. This is called the quantum back action of the
measurement, and the limit to sensitivity so imposed is referred to as
the standard quantum limit. One of the pioneers in this field has been
Braginsky in various studies of measurement aspects of the fluctuations
of light caused by the moving mirror~\cite{Braginsky75,Braginsky88}.

In real devices which have been constructed so far, the quantum back
action noise in the measurement record is usually small compared to
that arising from classical sources of noise. However, as the
sensitivity of such devices increases it is expected that we  will
eventually obtain displacement sensors that are quantum  limited. The
quantum back-action noise has not yet been seen experimentally for
macroscopic devices, so seeing it is a topic of current interest. Once
the standard quantum limit has been achieved, this will not be the end
of the story, however. Various authors have shown that it is possible
to use contractive states~\cite{bql1}, or squeezed light~\cite{pace,bql2},
to reduce the quantum back action and therefore  increase the
sensitivity of the measurement even further.

The interferometer we consider here for measuring position consists
essentially of a cavity where one of mirrors is free to move. This
system is also of interest from the point of view of cavity QED.
Usually cavity QED experiments require optical cavities where the
atomic excitations and photons in the optical modes become entangled.
The dynamics follows from the interplay between these quantum
variables. However, a challenging realm for cavity QED experiments
involves instead an {\em empty} cavity (that is, a cavity  containing
no atoms or optical media) where the photons in the cavity mode
interact with the motion of one of the cavity mirrors. In this scheme,
the position of at least one mirror in the optical resonator is a
dynamic variable. The coupling between the photons and the mirror
position is simply the radiation pressure that stems from the momentum
transfer of $2 \hbar k$ per one reflected photon with the wavenumber
$k$. It has been been shown that this system may be used to generate
sub-Poissonian light in the output from the 
cavity~\cite{Heidmann94,Fabre94,Pinard95}. The moving mirror alters 
the photon statistics by changing the optical path length in a way 
that is proportional to the instantaneous photon number inside the
cavity. This system may also be used to create highly non-classical
states of the cavity field, such as Schr\"{o}dinger
cats~\cite{TM,BJK}, and might even be used to create cat states of the
mirror~\cite{BJK}. In addition, it has been shown that such a
configuration may be used to perform QND measurements of the light
field~\cite{QND}, and to detect the decoherence of the mirror, a topic
of fundamental interest in quantum measurement theory~\cite{BJK2}. Due
to recent technological developments in optomechanics, this area is
now becoming experimentally accessible. Dorsel {\em et al.} have
realised optical bistability with this system \cite{Dorsel}, and other
experiments, particularly to probe the quantum noise, are now in
progress~\cite{Tittonen98,Paris99}.

In order to create displacements that are large enough to be observed,
one is tempted to use a mirror having a well-defined mechanical
resonance with a very high quality factor $Q$. Thus, even when excited
with weak white noise driven radiation pressure, the mirror can be
displaced by a detectable amount at the mechanical resonance frequency
$\nu$. For such a mirror to behave fully quantum mechanically one needs
to operate at very low temperatures since the thermal energy $kT$ very
easily exceeds $\hbar \nu$. For example, a $\nu/2 \pi=100$ kHz resonance
is already significantly excited  at 5 $\mu$K. However, it is not
necessary to reach the fully quantum domain to observe the quantum
back action. By simultaneously combining a high optical quality factor
(ie. by using a high-finesse cavity) and a specially designed low mass
mirror with very high mechanical quality factor one can at typical
cryogenic temperatures create conditions where the radiation pressure
fluctuations (which are the source of the quantum mechanical
back-action referred to earlier) exceed the effects caused by thermal
noise. In this paper we discuss considerations for detecting this
quantum back-action noise.

There are already a number of publications dealing with quantum noise
in optical position measurements. Our main purpose here is to extend
this literature in two ways which are important when considering the
detection of the quantum noise. The first is the inclusion of the
effects of experimental sources of noise, such as the classical laser
noise and the noise from intracavity losses. The second is to perform
a quantum treatment of phase-modulation detection, so that the results
may be compared with those for homodyne detection. While this method
of phase detection is often used in practice, it has not previously 
been given a quantum mechanical treatment, which we show is important 
because previous semiclassical treatments have underestimated the shot 
noise. In addition to these main objectives, we also show that the 
standard Brownian motion master equation is not adequate to describe 
the thermal damping of the mirror, but that the corrected Brownian motion
master equation derived by Diosi~\cite{Diosi} rectifies this problem.

In section~\ref{sec2} we describe the configuration of the system.
In section~\ref{sec3} we perform a quantum mechanical analysis of
phase modulation detection. In section~\ref{sec4} we solve the
linearised equations of motion for the cavity/mirror system, using a
non-standard Brownian motion master equation which is of the Lindblad
form~\cite{Diosi}. In section~\ref{sec5} we use this solution to
obtain the noise power spectral density (which we refer to simply as
the {\em spectrum}) for a measurement of the phase quadrature using
phase modulation detection. In the first part of this section we
discuss each of the contributions and their respective forms. Next we
compare the spectrum to that which results if the standard
(non-Lindblad) Brownian motion master equation is used to describe the
thermal damping of the mirror, and also to that which would have been
obtained using homodyne detection rather than phase modulation
detection. Finally we show how the error in a measurement of the
position of the mirror may be obtained easily from the spectrum. We
evaluate explicitly the contribution to this error from various noise
sources, and plot these as a function of the laser power.
Section~\ref{sec6} concludes.

\section{The System}\label{sec2}
The system under consideration consists of a coherently driven
optical cavity with a moving mirror which will be treated as a quantum
mechanical harmonic oscillator. The light driving the cavity reflects
off the moving mirror and therefore fluctuations in the position
of the mirror register as fluctuations in the light output from the
cavity. In the limit in which the cavity damping rate is much larger
than the rate of the dynamics of the mirror (characterised by the
frequency of oscillation $\nu$ and the thermal damping rate $\Gamma$)
the phase fluctuations of the output light are highly correlated with
the fluctuations of the position of the mirror and constitute a
continuous position measurement of the mirror~\cite{MJW}.

An experimental realisation will therefore involve a continuous
phase-quadrature measurement of the light output from the cavity to
determine the output spectrum of the phase-quadrature fluctuations.
The nature of the detection scheme used to measure the phase
quadrature is of interest to us, as we shall see that it will effect
the relationship of the shot noise to the other noise sources in the
measured signal. Quantum theoretical treatments usually assume the
use of homodyne detection~\cite{MJW,Heidmann94,Fabre94,Pinard95}.
However this is often not used in practice~\cite{HH,HSKO}. Many current
experiments use instead {\em phase 

\begin{figure}
\leavevmode
\epsfxsize=8cm
\epsfbox{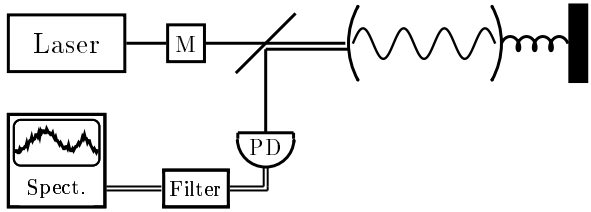}
\caption{\narrowtext The light output from the laser is phase-modulated
at frequency $\Delta$ in the modulator M, and from there drives the
cavity. The front mirror of the cavity is fixed, while the back mirror
is a mechanical harmonic oscillator. The diagonal line is a shorthand
representation for the arrangement which isolates the laser from the
cavity output, and the light that is reflected back off the front
mirror. All this light falls upon a photo-detector PD, and the
photo-detection signal is demodulated (to pick out the phase quadrature
signal) before going to a spectrum analyser.}
\label{theoryfig}
\end{figure}

$\!\!\!\!\!\!\!\!$ modulation} 
detection~\cite{Tittonen98}, which was developed by
Bjorklund~\cite{Bjorklund1,Bj2} in 1979. Before we treat the dynamics
of the cavity field/oscillating mirror system, to determine the effect
of various noise sources, we will spend some time in the next section
performing a quantum mechanical treatment of phase modulation
detection. We will focus on this scheme throughout our treatment, and
compare the results with those for Homodyne detection. A diagram of the
experimental arrangement complete enough for the theoretical analysis
is given in figure~\ref{theoryfig}. We note that in practice a feedback
scheme is used to lock the laser to the cavity so as to stabilise the
laser frequency. For an analysis of the method and an expression for
the resulting classical phase noise the reader is referred to
references~\cite{HH} and~\cite{HSKO}. We do not need to treat this
feedback explicitly, however. Its effect may be taken into account by
setting the value of the classical laser phase noise
in our analysis to the level it provides.

\section{Phase Modulation Detection}\label{sec3}
The laser which drives the cavity is isolated from the cavity output,
and the entirety of this output falls upon a photo-detector. In order
that the photo-detection signal contain information regarding the
phase quadrature, the laser field is modulated at a frequency $\Delta$,
which is chosen to be much greater than the natural frequency of the
harmonic mirror. The sidebands that result from this modulation are
chosen to lie far enough off-resonance with the cavity mode that they do
not enter the cavity and are simply reflected from the front mirror. From
there they fall upon the photo-detector. The result of this is that the
output phase quadrature signal appears in the photodetection signal as a
modulation of the amplitude of a `carrier' at frequency $\Delta$. This
is then demodulated (by multiplying by a sine wave at the modulation
frequency and time averaging) to pick out the phase quadrature signal,
and from there the spectrum may be calculated.

First consider the laser output field, which is essentially classical;
it is a coherent state in which the amplitude and phase are not
completely stable and therefore contain some noise. This means that
the field from the laser actually contains frequencies in a small
range about its central frequency. The laser field may therefore be
described by a set of coherent states with frequencies in this range.
As a result it is possible to perform a unitary transformation on the
mode operators such that the amplitude of each of the coherent states
is replaced by a complex number, and the quantum state of the field is
simply the vacuum~\cite{Gardiner2,Mollow,HMWPhD}. This separates out
the classical variations in the field from the quantum contribution,
and allows us to write the output from the laser as
\begin{equation}
  \beta + \delta a_{\mbox{\scriptsize in}}(t) + \delta x(t) + i\delta
y(t)
\end{equation}
In this expression $\beta$ is the average coherent amplitude of the
field, which we choose to be real, and $\beta^2$ is the photon flux. The
deviations from this average are given by $\delta x(t)$, being the
classical amplitude noise, and $\delta y(t)$, being the classical phase
noise. The quantum noise, which may be interpreted as arising from the
vacuum quantum field, is captured by the correlation function of the
field operator
$\delta a_{\mbox{\scriptsize in}}(t)$. Here the subscript refers
to the field's relation to the cavity, and not the laser. The
correlation functions of the various noise sources are
\begin{eqnarray}
  & &  \langle \delta a_{\mbox{\scriptsize in}}
(t)\delta a_{\mbox{\scriptsize in}}^\dagger(t+\tau)\rangle
=\delta(\tau) \nonumber \\
  & &  \langle \delta a_{\mbox{\scriptsize in}}^\dagger(t)
\delta a_{\mbox{\scriptsize in}}(t+\tau)\rangle = 0
\nonumber \\
  & &  \langle \delta x(t)\delta x(t+\tau)
\rangle = G_{\mbox{\scriptsize x}}(\tau) \nonumber \\
  & &  \langle \delta y(t)\delta y(t+\tau) \rangle =
G_{\mbox{\scriptsize y}}(\tau) , \nonumber
\end{eqnarray}
where we have left the classical noise sources arbitrary. This allows
them to be tailored to describe the output from any real laser source
at a later time. However, we will assume that the modulation frequency,
$\Delta$, is choosen large enough so that the classical noise is negligible
at this frequency. This is what is done in practice. The average values of the
three noise sources, $\delta a_{\mbox{\scriptsize in}}$, $\delta x$ and
$\delta y$ are zero, as are all the cross correlations.

Before the laser field enters the cavity, it passes through a phase
modulator. This is a classical device which modulates the phase of the
coherent amplitude of the beam, and as such leaves the quantum noise
unaffected. The phase is modulated sinusoidally, the result of which is
to transform the time dependent coherent amplitude, given by
$\beta + \delta x(t) + i\delta y(t)$, into~\cite{Bjorklund1,Yariv}
\begin{equation}
  (\beta + \delta x(t) + i\delta y(t))\sum_{n=-\infty}^{n=\infty} J_n(M)
e^{in\Delta t}
\end{equation}
where $J_n$ is the $\mbox{n}^{\mbox{\scriptsize \em th}}$ Bessel
function, $\Delta$ is the frequency of the modulation, and $M$ is
referred to as the modulation index, being determined by
the amplitude of the sinusoidal modulation. For phase modulation
detection, the modulation index is typically chosen to be much less than
unity so that $J_0\approx 1$, $J_{\pm 1}=\pm M/2 \equiv \pm \varepsilon$,
$\varepsilon \ll 1$, and all other terms vanish. The laser field after
modulation is then
\begin{equation}
   (\beta + \delta x(t) + i\delta y(t))(1 + \varepsilon e^{-i\Delta t} -
\varepsilon e^{i\Delta t}) + \delta a_{\mbox{\scriptsize in}}(t) .
\end{equation}
Using now the input-output relations of Collet and
Gardiner~\cite{inout},
the field output from the cavity is
\begin{eqnarray}
   a_{\mbox{\scriptsize out}}(t) & = & - (\beta + \delta x(t) +
i\delta y(t))
(1 + \varepsilon e^{-i\Delta t} - \varepsilon e^{i\Delta t})
\nonumber \\
                                 &   & - \delta a_{\mbox{\scriptsize
in}}(t) +
\sqrt{\gamma} (\delta a(t) + \alpha),
\end{eqnarray}
in which $a(t)=\delta a(t) + \alpha$ is the operator describing the
cavity mode, and $\gamma$ is the decay constant of the cavity due to
the input coupling mirror. We are interested in the steady state
behaviour, and we choose $\alpha$ to be the average steady state
field strength in the cavity. In addition, in order to solve the
equations of motion for the cavity we will linearise the system about
the steady state, which requires that $\langle \delta a^{\dagger}(t)
\delta a(t) \rangle \ll |\alpha|^{2}$. The operator describing the
photo-current from the photo-detector is
\end{multicols}
\vspace{-0.5cm}
\noindent\rule{0.5\textwidth}{0.4pt}\rule{0.4pt}{\baselineskip}
\widetext
\begin{eqnarray}
  I(t) & = & a_{\mbox{\scriptsize out}}(t)^\dagger
a_{\mbox{\scriptsize out}}(t)
\nonumber \\
& = & \tilde{\alpha}^2 + \tilde{\alpha}(\delta X_{\mbox{\scriptsize
out}} - 2\,\delta x) +
 2\varepsilon\beta\sin(\Delta t)\left(\delta Y_{\mbox{\scriptsize
out}} - \frac{2\sqrt{\gamma}\alpha\,\delta y}{\beta}\right) + (2\varepsilon\beta)^2\sin^2(\Delta
t)\left(1+\frac{2\,\delta x}{\beta}\right)
\end{eqnarray}
\begin{multicols}{2}
in which
\begin{eqnarray}
\delta X_{\mbox{\scriptsize out}}(t) & = & \sqrt{\gamma}\,\delta X(t) -
\delta X_{\mbox{\scriptsize in}}(t) \nonumber \\
\delta Y_{\mbox{\scriptsize out}}(t) & = & \sqrt{\gamma}\,\delta Y(t)
-
\delta Y_{\mbox{\scriptsize in}}(t) \nonumber
\end{eqnarray}
and
\begin{eqnarray}
\delta X(t) & = & \delta a(t) + \delta a^\dagger(t) \nonumber \\
  \delta Y(t) & = & -i (\delta a(t) - \delta a^\dagger(t)) \nonumber
\\
  \delta X_{\mbox{\scriptsize in}}(t) & = & \delta
a_{\mbox{\scriptsize in}}(t) +
\delta a^\dagger_{\mbox{\scriptsize in}}(t) \nonumber \\
  \delta Y_{\mbox{\scriptsize in}}(t) & = & -i (\delta
a_{\mbox{\scriptsize in}}(t) -
\delta a^\dagger_{\mbox{\scriptsize in}}(t)).
\end{eqnarray}
We have also set $\tilde{\alpha}=(-\beta+\sqrt{\gamma}\alpha)$, and
assumed this to be real. To obtain the phase quadrature signal we
demodulate, which involves multiplication by a sine wave at frequency
$\Delta$, and subsequent averaging over a time, $T$. This time must
 be long compared to $1/\Delta$, but short compared to the time scale
of the phase quadrature fluctuations. The signal is therefore given by
\begin{eqnarray}
 R(t) & = &  \frac{1}{T}\int_0^T \sin(\Delta \tau) I(t+\tau)\;
\mbox{d}\tau ,
\end{eqnarray}
We now must evaluate this to obtain $R(t)$ explicitly in terms of the
phase quadrature. Writing out the integral, and dropping everything
which averages to zero (that is, which is not passed by the low-pass
filtering) we obtain
\begin{eqnarray}
 R(t) & = & \left( \frac{\varepsilon\beta}{T} \right) \left[ \int_0^T
\delta Y_{\mbox{\scriptsize out}}(t+\tau) d\tau \right] \nonumber \\
      & - & \left(\frac{2\varepsilon\sqrt{\gamma}\alpha}{T}\right)
\left[
\int_0^T \delta y(t+\tau) \; d\tau \right] \nonumber \\
      & + & q_1(t) + q_2(t) ,
\end{eqnarray}
where
\end{multicols}
\vspace{-0.5cm}
\noindent\rule{0.5\textwidth}{0.4pt}\rule{0.4pt}{\baselineskip}
\widetext
\begin{eqnarray}
 q_1(t) & = & - \left( \frac{\beta-\sqrt{\gamma}\alpha}{T} \right)
\mbox{Re}\left[ \int_0^T ie^{-i\Delta(t+\tau)}\,\delta X_{\mbox{\scriptsize
in}}(t+\tau)  \; d\tau \right], \\
 q_2(t) & = &  \left( \frac{\varepsilon\beta}{T} \right) \mbox{Re}\left[
\int_0^T e^{-i2\Delta(t+\tau)}\,\delta Y_{\mbox{\scriptsize in}}(t+\tau)
\; d\tau \right] .\label{defq2}
\end{eqnarray}
\begin{multicols}{2}
In deriving this expression we have assumed that the classical
laser  noise is  
Note that we choose $T$ to be much smaller than the time-scale upon 
which $\delta Y$ and $\delta y$ change, so that the integration is
essentially equivalent to multiplication by $T$, an effect which is
canceled by the division by $T$. However, we should also note that
$\delta Y_{\mbox{\scriptsize out}}$ contains $\delta
Y_{\mbox{\scriptsize in}}$, so that in replacing the first term in
$R(t)$ by $\varepsilon\beta \; \delta Y_{\mbox{\scriptsize out}}(t)$
we must remember that this only contains the frequency components of
$\delta Y_{\mbox{\scriptsize in}}$ in a bandwidth of $1/T$ around zero
frequency. The result of this is that $\delta Y_{\mbox{\scriptsize
out}}(t)$ is uncorrelated with $q_1$ and $q_2$, being the quantum noise
in the bandwidth $1/T$ around the frequencies $\Delta$ and $2\Delta$
respectively. We need to know the correlation functions of these noise
sources, and whether or not they are correlated with any of the other
terms in $R(t)$. It is clear that $q_1$ and $q_2$ are not correlated
over separation times greater than $2T$. Using Eq.(\ref{defq2}) to
evaluate the correlation function of $q_2$, for example, we have
\begin{equation}
  \langle q_2(t)q_2(t+\tau)\rangle = \left\{ \begin{array}{cl}

\frac{(\varepsilon\beta)^2}{2} \left( \frac{T-|\tau|}{T^2} \right) &
\mbox{for} \; |\tau| \leq T \\
                                              0 & \mbox{otherwise.}
                                            \end{array}  \right.
\end{equation}
On the time-scale of the fluctuations of $\delta Y$ we can approximate
this as a delta function, so that $q_1$ and $q_2$ (and also $\delta
Y_{\mbox{\scriptsize in}}$) are still effectively white noise sources.
We may therefore write
\begin{eqnarray}
 R(t) & = & \varepsilon\beta \; \delta Y_{\mbox{\scriptsize out}}(t)
+ q_1(t) + q_2(t) - 2\varepsilon\sqrt{\gamma}\alpha\delta y(t) ,
\label{signal}
\end{eqnarray}
and the correlation functions of $q_1$ and $q_2$ are
\begin{eqnarray}
  & & \langle q_1(t)q_1(t+\tau)\rangle = (1/2)(\beta -
\sqrt{\gamma}\alpha)^2 \; \delta(\tau) ,\nonumber \\
  & & \langle q_2(t)q_2(t+\tau)\rangle = (1/2)(\varepsilon\beta)^2 \;
\delta(\tau) .
\end{eqnarray}
The signal therefore contains the phase quadrature of the output
field, $\delta Y_{\mbox{\scriptsize out}}(t)$, plus three noise terms.
While the last term, being the input classical phase noise, is
correlated with $\delta Y_{\mbox{\scriptsize out}}(t)$, $q_1$ and
$q_2$ are not. Taking the Fourier transform of the signal,
\begin{equation}
  R(\omega) = \frac{1}{\sqrt{2\pi}}\int_{-\infty}^{\infty}
\!\!\!\!\!\! R(t) e^{-i\omega t} \; dt ,
\end{equation}
we may write
\begin{equation}
  R(\omega) = \varepsilon\beta \; \delta Y_{\mbox{\scriptsize
out}}(\omega) +
\sum_{i=1}^2q_j(\omega) - 2\varepsilon\sqrt{\gamma}\alpha\,\delta
y(\omega) .
\label{eq15}
\end{equation}
This is the Fourier transform of the signal in the case of phase
modulation detection. If we were to use ideal homodyne detection this
would be instead~\cite{HMWhom}
\begin{equation} \label{above}
  R_{\mbox{\scriptsize h}}(\omega) =
  \kappa \tilde{\beta}( \delta Y_{\mbox{\scriptsize out}}(\omega)
- 2 \,\delta y(\omega)),
\end{equation}
where $\tilde{\beta}$ is the amplitude of the local oscillator and
$\kappa$ is the reflectivity of the beam splitter used in the homodyne
scheme. Thus, in the case of phase modulation detection, there are
two white noise sources which do not appear in homodyne detection.
They stem from the fact that the phase quadrature detection method
is demodulating to obtain a signal at a carrier frequency. Because
the quantum noise is broad band (in particular, it is broad
compared to the carrier frequency) the demodulation picks up the
quantum noise at $\Delta$ and $2\Delta$. There is also a term
from the classical phase noise in the sidebands. We note that the 
contribution from the quantum noise at $2\Delta$ has been omitted 
from previous semiclassical treatments, with the result that the 
shot noise has been underestimated by 
$(1/2)(\varepsilon\beta)^2$~\cite{semicl}.
For unbalanced homodyne detection there will also be an extra
contribution from the noise on the local oscillator, which may be
suppressed (in the limit of an intense local oscillator) with the use
of balanced homodyne detection~\cite{YS}.

Returning to Eq.(\ref{eq15}) for the demodulated signal, the next step
is to solve the equations of motion for the system operators to obtain
$\delta Y(\omega)$ in terms of the input noise sources. We can then
readily calculate $\langle R(\omega)R(\omega')\rangle$, which appears in
the form
\begin{equation}
\langle R(\omega)R(\omega')\rangle = S(\omega)\delta(\omega+\omega').
\end{equation}
The delta function in $\omega$ and $\omega'$ is a result of the
stationarity of $R(t)$, and $S(\omega)$ is the {\em power spectral
density}, which we will  refer to from now on simply as the {\em
spectrum}. This is useful because, when  divided by $2\pi$, it gives
the average  of the square of the signal per unit frequency (The
square of the  signal is universally referred to as the {\em power},
hence the name power spectral  density). Since the noise has zero
mean, the square average is the variance,  and thus the spectrum
provides us with information regarding the  error in the signal due to
the noise. The spectrum is  also a Fourier transform of the
autocorrelation function~\cite{Gardiner}. The specific relation, using
the definitions we have introduced above, is
\begin{equation}
  S(\omega) = \int_{-\infty}^{\infty} \!\!\!\! \langle
R(0)R(\tau)\rangle e^{-i\omega\tau} d \tau,
\end{equation}
and as the autocorrelation function has units of $\mbox{s}^{-2}$, the
spectrum has units of $\mbox{s}^{-1}$. To determine the spectrum
experimentally the phase of the signal is measured for a time long
compared to the width of the auto-correlation function, and the Fourier
transform is taken of the result. Taking the square modulus of this
Fourier transform, and dividing by the duration of the measurement
obtains a good approximation to the theoretical spectrum. We proceed
now to calculate this spectrum.

\section{The Dynamics of the System}\label{sec4}
Excluding coupling to reservoirs, the Hamiltonian for the combined
system of the  cavity mode and the mirror is~\cite{JTCW}
\begin{eqnarray}
 H & = & \hbar\omega_0 a^{\dagger}a + \frac{p^2}{2m} +
\frac{1}{2}m\nu^2 q^2 - \hbar ga^\dagger aq  \nonumber \\
   &   &  + \hbar
\left\{ i\left[ E + \sqrt{\gamma}\,\delta x(t) + i\sqrt{\gamma}\,\delta
y(t)\right]
a^\dagger + \mbox{H.c.}\right\} .
\label{Ham}
\end{eqnarray}
In this equation $\omega_0$ is the frequency of the cavity mode, $q$
and $p$ are the position and momentum operators for the mirror
respectively, $m$ and $\nu$ are the mass and angular frequency of the
mirror, $g = \omega_{0}/L$ is the coupling constant between the cavity
mode and the mirror (where $L$ is the cavity length), and $a$ is the
annihilation operator for the mode. The classical driving of the cavity
by the coherent input field is given by $E$ which has dimensions
of $\mbox{s}^{-1}$, and is related to the input laser power $P$ by
 $E=\sqrt{P\gamma/(\hbar\omega_0)}=\sqrt{\gamma}\beta$.
The classical laser noise appears as noise on this driving term.

The moving mirror is a macroscopic object at temperature $T$, and as
such is subject to thermal noise. While it is still common to use the
standard Brownian motion master equation (SBMME)~\cite{Gardiner,SBMME}
to model such noise, as it works well in many situations, it turns out
that it is not adequate for our purposes. This is because
it generates a clearly non-sensical term in the spectrum. As far as we
know this is the first time that it has been demonstrated to fail in
the steady state. Discussions regarding the SBMME, and non-Lindblad
master equations may found in references~\cite{MGetc,Haake}.
We will return to this point once we have calculated the spectrum. We
use instead the corrected Brownian motion master equation (CBMME)
derived by Diosi~\cite{Diosi}, to describe the thermal damping
of the mirror, as this corrects the problems of the SBMME. In
particular, we use the CBMME in which the cut-off frequency of the
thermal reservoir is assumed to be much smaller than
$k_{\mbox{\scriptsize B}} T/\hbar$. For current experiments
$k_{\mbox{\scriptsize B}} T/\hbar$ is greater than 10 GHz, so this
assumption appears reasonable, and leads to the simplest Lindblad-form
Brownian motion master equation. Using this CBMME, and the standard
master equation for the cavity losses (both internal and external), the
quantum Langevin equations of motion for the system are given by
\begin{eqnarray} \label{qle1}
\dot{a} & = & -\frac{i}{\hbar}[a,H] -
\left(\frac{\gamma+\mu}{2}\right)a +
\sqrt{\gamma}\,\delta a_{\mbox{\scriptsize in}}(t) +
\sqrt{\mu}b_{\mbox{\scriptsize in}}(t)  \\
\dot{q} & = & -\frac{i}{\hbar}[q,H] +
\hbar(\Gamma/6mkT)^{\frac{1}{2}}\eta(t)\label{lang2}\\
 \dot{p} & = & -\frac{i}{\hbar}[p,H] - \Gamma p +
(2 m \Gamma kT)^{\frac{1}{2}} \xi(t) , \label{lang3}
\label{qle3}
\end{eqnarray}
in which the correlation functions for the Brownian noise sources are
\begin{eqnarray}
\langle \xi(t) \xi(t') \rangle &=& \delta(t-t'), \\
\langle \eta(t) \eta(t') \rangle &=& \delta(t-t'), \\
  \langle \xi(t) \eta(t') \rangle &=& -i(\sqrt{3}/2)\delta(t-t'),  \\
\langle \eta(t) \xi(t') \rangle &=& i(\sqrt{3}/2)\delta(t-t') .
\end{eqnarray}
In these equations $\gamma$ is the decay constant describing
transmission through the input coupling mirror. All `internal' cavity
losses including absorption, scattering and loss through the movable
mirror are included separately via the decay constant $\mu$, and the
corresponding vacuum fluctuations via the operator
$b_{\mbox{\scriptsize in}}(t)$. The effect of mechanical
damping and thermal fluctuations of the mirror are given by the noise
sources $\xi(t)$ and $\eta(t)$ and the mechanical damping constant
$\Gamma$.

We note here that if we were to use the standard Brownian motion
master equation~\cite{SBMME,Gardiner}, Eqs.(\ref{lang2}) and
(\ref{lang3}) would instead be given by
\begin{eqnarray}
\dot{q} & = & -\frac{i}{\hbar}[q,H]  \\
\dot{p} & = & -\frac{i}{\hbar}[p,H] - \Gamma p +
(2 m \Gamma kT)^{\frac{1}{2}} \zeta(t) .
\end{eqnarray}
where $\langle\zeta(t)\zeta(t')\rangle = \delta(t-t')$. These
Langevin equations do not preserve the commutation relations of the
quantum mechanical operators, and as a result it is clear that the
description cannot be entirely correct.

Calculating the commutators in Eqs.(\ref{qle1}) to (\ref{qle3}), we
obtain
\end{multicols}
\vspace{-0.5cm}
\noindent\rule{0.5\textwidth}{0.4pt}\rule{0.4pt}{\baselineskip}
\widetext
\begin{eqnarray}
 \dot{a} & = & E -\left(\frac{\gamma+\mu}{2}\right)a + igaq +
\sqrt{\gamma}\,\delta a_{\mbox{\scriptsize in}}(t) +
\sqrt{\mu}b_{\mbox{\scriptsize in}}(t) +
\sqrt{\gamma}\,\delta x(t) + i\sqrt{\gamma}\,\delta y(t) \\
 \dot{q} & = & \frac{p}{m} + \hbar(\Gamma/6mkT)^{\frac{1}{2}}
\eta(t) \\
 \dot{p} & = & -m\nu^2 q + \hbar g a^\dagger a - \Gamma p +
(2 m \Gamma kT)^{\frac{1}{2}} \xi(t) .
\end{eqnarray}
\begin{multicols}{2}
Introducing a cavity detuning $\delta\omega$ (that is, setting the
cavity resonance frequency in the absence of any cavity field to
$\omega_{\mbox{\scriptsize c}}= \omega_0 + \delta\omega$), and solving
these equations for the steady state average values we obtain
\begin{eqnarray}
  \langle a\rangle_{\mbox{\scriptsize ss}} & = &
\frac{2E}{\gamma+\mu} \equiv \alpha  \\
  \langle q\rangle_{\mbox{\scriptsize ss}} & = & \frac{\hbar
g}{m\nu^2}|\alpha|^2 \\
  \langle p\rangle_{\mbox{\scriptsize ss}} & = & 0
\end{eqnarray}
where we have set the detuning to $\delta\omega=g\langle
x\rangle_{\mbox{\scriptsize ss}}$ to bring the cavity on resonance with
the driving field in the steady state. Linearising the quantum Langevin
equations about the steady state values, and writing the result in
terms of the field quadratures, we obtain the following linear
equations
\end{multicols}
\vspace{-0.5cm}
\noindent\rule{0.5\textwidth}{0.4pt}\rule{0.4pt}{\baselineskip}
\begin{equation}
   \left(  \begin{array}{c}
                            \delta\dot{X}  \\
                            \delta\dot{Y}  \\
                            \delta\dot{Q}  \\
                            \delta\dot{P}
           \end{array} \right)  =
   \left(  \begin{array}{cccc}
  -\frac{\gamma+\mu}{2}   & 0                       &  0 & 0  \\
  0      & -\frac{\gamma+\mu}{2}   &  \chi\alpha   & 0   \\
  0       & 0                       &  0            & \nu  \\
  \chi\alpha   & 0             & -\nu       & -\Gamma
           \end{array} \right)
   \left(  \begin{array}{c}
                         \delta X  \\
                         \delta Y  \\
                         \delta Q  \\
                         \delta P
           \end{array} \right) +
   \left(  \begin{array}{c}
\sqrt{\gamma}\,\delta X_{\mbox{\scriptsize in}}(t) +
\sqrt{\mu}\,\delta X_{\mbox{\scriptsize b,in}}(t) +
2\sqrt{\gamma}\,\delta x(t) \\
\sqrt{\gamma}\,\delta Y_{\mbox{\scriptsize in}}(t) +
\sqrt{\mu}\,\delta Y_{\mbox{\scriptsize b,in}}(t) +
2\sqrt{\gamma}\,\delta y(t)            \\
(\Gamma\hbar\nu/3kT)^{\frac{1}{2}}  \eta(t)  \\
(4\Gamma kT/(\hbar\nu))^{\frac{1}{2}} \xi(t)
           \end{array} \right) .
\label{dyn}
\end{equation}
\begin{multicols}{2}
In this set of equations we have scaled the position and momentum
variables using
\begin{eqnarray}
   \delta Q & = & \sqrt{\frac{2m\nu}{\hbar}}(q - \langle
q\rangle_{\mbox{\scriptsize ss}}) \\
   \delta P & = & \sqrt{\frac{2}{m\hbar\nu}}(p - \langle
p\rangle_{\mbox{\scriptsize ss}}) .
\end{eqnarray}
and we have defined $\chi\equiv g(2\hbar/m\nu)^{1/2}$, which has
units of s$^{-1}$. The quadratures for the input noise due to
intracavity losses are given by
\begin{eqnarray}
   \delta X_{\mbox{\scriptsize b,in}} & = & b_{\mbox{\scriptsize in}}
+ b_{\mbox{\scriptsize in}}^\dagger \\
   \delta Y_{\mbox{\scriptsize b,in}} & = & -i(b_{\mbox{\scriptsize
in}}
- b_{\mbox{\scriptsize in}}^\dagger) .
\end{eqnarray}
Without loss of generality we have chosen the input field amplitude
to be real ($\mbox{Im}[\beta]=0$), so that the input phase quadrature
is given by $Y_{\mbox{\scriptsize in}}$. We now solve the dynamics
(\ref{dyn}) in the frequency domain in order to obtain the spectrum
directly from the solution. To switch to the frequency domain we Fourier
transform all operators and noise sources. In particular we have, for
example
\begin{eqnarray}
  \delta a(\omega) & \equiv &
\frac{1}{\sqrt{2\pi}}\int_{-\infty}^{\infty}
 \!\!\!\!\!\! \delta a(t) e^{i\omega t} \; dt , \\
  \delta a^\dagger(\omega) & \equiv &
\frac{1}{\sqrt{2\pi}}\int_{-\infty}^{\infty}
 \!\!\!\!\!\! \delta a^\dagger(t) e^{i\omega t} \; dt = [\delta
a(-\omega)]^\dagger .
\end{eqnarray}
Rearranging the transformed equations, the solution is given by
\begin{equation}
 (\delta X(\omega), \delta Y(\omega), \delta Q(\omega), \delta
P(\omega))^T =
M(\omega) {\bf n}(\omega)
\end{equation}
where ${\bf n}(\omega)$ is the vector of transformed noise sources.
If we write the matrix elements of $M(\omega)$ as
$M_{ij}(\omega)=m_{ij} (\omega)/D(\omega)$, then
\begin{equation}
  D(\omega) = \left((\gamma+\mu)/2 - i\omega \right)^2 \left( \nu^2 -
\omega^2 -
i\Gamma\omega \right)
\end{equation}
and the non-zero $m_{ij}$ are given by
\begin{eqnarray}
  m_{11} & = & \left((\gamma+\mu)/2 - i\omega \right)  \left( \nu^2 -
\omega^2 -
i\Gamma\omega \right) = m_{22}, \nonumber \\
  m_{21} & = & \chi^2\alpha^2\nu \nonumber \\
  m_{23} & = & \chi\alpha (\Gamma - i\omega)\left((\gamma+\mu)/2 -
i\omega \right) , \nonumber \\
  m_{24} & = & \chi\alpha\nu \left((\gamma+\mu)/2 - i\omega \right) =
m_{31} , \nonumber \\
  m_{33} & = & (\Gamma - i\omega)\left((\gamma+\mu)/2 - i\omega
\right)^2 , \nonumber \\
  m_{34} & = & \nu\left((\gamma+\mu)/2 - i\omega \right)^2 = -m_{43},
\nonumber \\
  m_{41} & = & -i\chi\alpha\omega \left((\gamma+\mu)/2 - i\omega
\right) , \nonumber \\
  m_{44} & = & -i\omega \left((\gamma+\mu)/2 - i\omega \right)^2  .
\end{eqnarray}
We have now solved the equations of motion for the system in
frequency space. The spectra of the system variables may now be
calculated in terms of the input noise sources. Using the
input-output relations, which give the output field in terms of the
system variables and the input noise sources, the spectra of the
output field, and hence of the measured signal, may be obtained. Note
that quantum mechanics plays no role in the solution of the motion of
the system. The linear equations of motion may as well be equations
for classical variables. The only part that quantum mechanics plays
in determining the spectra of the system variables is that some of
the input noise sources are quantum mechanical. That is, their
correlation functions are determined by quantum mechanics. In fact,
if all the noise sources had purely classical correlation functions,
then the SBMME Langevin equations would not lead to any problems, as
they are perfectly correct as equations of motion for a classical
system.

\section{The Power Spectral Density}\label{sec5}
To calculate the spectrum of the signal, we require the correlation
functions of the input noise sources. To reiterate, these are
\begin{eqnarray}
  \langle \delta X_{\mbox{\scriptsize in}}(\omega)
\delta X_{\mbox{\scriptsize in}}(\omega')\rangle & = & \langle \delta
Y_{\mbox{\scriptsize in}}(\omega)\delta Y_{\mbox{\scriptsize in}}
(\omega')\rangle = \delta(\omega+\omega'), \nonumber \\
   \langle \delta X_{\mbox{\scriptsize in}}(\omega)
\delta Y_{\mbox{\scriptsize in}}(\omega')\rangle & = & - \langle
 \delta Y_{\mbox{\scriptsize in}}(\omega)\delta X_{\mbox{\scriptsize
in}}
(\omega')\rangle = i\delta(\omega+\omega'), \nonumber
\end{eqnarray}
and similarly for $\delta X_{\mbox{\scriptsize b,in}}(\omega)$ and
$\delta Y_{\mbox{\scriptsize b,in}}(\omega)$. The correlation
functions
 for the classical laser noise, and thermal noise sources are
\begin{eqnarray}
  \langle \delta x(\omega)\delta x(\omega')\rangle & = &
\tilde{G}_x(\omega)\delta(\omega+\omega'), \nonumber \\
  \langle \delta y(\omega)\delta y(\omega')\rangle & = &
\tilde{G}_y(\omega)\delta(\omega+\omega'), \nonumber \\
  \langle \xi(t) \xi(t') \rangle &=& \langle \eta(t) \eta(t') \rangle
= \delta(t-t'), \nonumber \\
  \langle \eta(t) \xi(t') \rangle &=&  -\langle \xi(t) \eta(t')
\rangle = i(\sqrt{3}/2)\delta(t-t').
\end{eqnarray}
After some calculation we obtain the spectrum of the signal for phase
modulation detection as
\end{multicols}
\vspace{-0.5cm}
\noindent\rule{0.5\textwidth}{0.4pt}\rule{0.4pt}{\baselineskip}
\begin{eqnarray}
  \frac{1}{(\varepsilon\beta)^2}S(\omega)
     & = &   \frac{1}{2} \left[ 3 + \left(  \frac{\gamma-\mu}
{\varepsilon(\gamma+\mu)} \right)^2 \right]
           + \gamma (\gamma + \mu + 4\gamma\tilde{G}_x(\omega))
\left[\frac{(\chi^2\alpha^2 \nu)^2}{|D(\omega)|^2}\right] \nonumber \\
     &   & + \mbox{\normalsize $4\tilde{G}_y(\omega)$}
             \left[ \frac{4\gamma^2}{(\gamma+\mu)^2}\left( \frac{\omega^2}
{\left( \frac{\gamma+\mu}{2} \right)^2 + \omega^2 } \right)
\right] \nonumber
\\
     &   & + \gamma (\chi\alpha)^2 \Gamma \left( 4\nu^2 T_{\mbox{s}}
+
\frac{1}{3}(\Gamma^2 + \omega^2) T_{\mbox{s}}^{-1} \right)
\left[ \frac{\left( \frac{\gamma+\mu}{2} \right)^2 + \omega^2}
{|D(\omega)|^2} \right] ,
\label{eqpsd}
\end{eqnarray}
\begin{multicols}{2}
where
\begin{equation}
  |D(\omega)|^2 = \left[(\gamma+\mu)^2/4  + \omega^2\right]^2
\left[(\nu^2 - \omega^2)^2 +
 \Gamma^2\omega^2\right],
\end{equation}
and $T_{\mbox{s}}$ is a dimensionless scaled temperature given
by $T_{\mbox{s}}=(k_{\mbox{\scriptsize B}}/(\hbar\nu))T$.
This phase-fluctuation spectrum may be thought of as arising
in the following way. The mechanical harmonic oscillator,
which is the moving mirror, is driven by  various noise sources,
both quantum mechanical and classical in origin, and the resulting
position fluctuations of the mirror are seen as fluctuations
in the phase of the light output from the cavity.

Let us examine the origin of the various terms in the spectrum in
turn. The first two terms, which appear in the first set of square
brackets, are independent of the frequency, and are the contribution
from the (quantum mechanical) shot noise of the light. The first term
has the factor of three (rather than a factor of two which would be the
case for homodyne detection) due to the contribution from $q_2(t)$. The
second term is the contribution from $q_1(t)$.

The next three terms, which multiply the second set of square brackets,
are the back-action of the light on the position of the mirror, noise
from internal cavity losses, and the classical amplitude noise on the
laser, respectively. Note that the only distinction between the
back-action and the internal losses is that former is proportional to
the loss rate due to the front mirror, and the latter is proportional
to the internal loss rate. It is easily seen that these noise sources
should have the same effect upon the position of the mirror: the
back-action is due to the random way in which photons bounce off the
mirror, whereas the internal losses are due to the similarly random
way in which photons are absorbed by the mirror, (or anything else in
the cavity). The amplitude fluctuations of the laser also affect the
mirror in the same manner, but since these fluctuations are not
white noise (as is the case with the quantum noise which comes from the
photon `collisions'), the response function of the mirror is multiplied
by the spectrum of the amplitude fluctuations.

The term which appears in the third set of square brackets is due to
the classical phase fluctuations of the laser. 
Clearly this has quite a different form from that due to the quantum
noise and the classical amplitude fluctuations. In particular, it is
not dependent upon the coupling constant, $g$, because it is derived
more or less directly from the input phase noise. Conversely, the noise
that derives from the amplitude fluctuations has its origin from the
fact that the amplitude fluctuations first drive the mirror, and it is
the resulting position fluctuations which cause the phase fluctuations
in the output.  The classical phase noise term includes a contribution
from the laser phase noise reflected from the cavity (that is, the
term given explicitly in Eq.(\ref{eq15})), and a contribution from the
phase noise on the light which has passed through the cavity
(being a part of $Y_{\mbox{\scriptsize out}}$). 

The final two terms, which multiply the fourth set of square brackets,
are due to the thermal fluctuations of the mirror. Note that these terms
are only valid in the region in which $k_{\mbox{\scriptsize B}} T \gg
\hbar\nu$.

Finally we note that we do not see squeezing in the spectrum of phase
quadrature fluctuations. This is because squeezing is produced when
the cavity detuning is chosen so that the steady state detuning is
non-zero~\cite{Fabre94}. We have chosen to set the steady state
detuning to zero in this treatment as we are not concerned here
with reducing the quantum noise.

In what follows we examine various aspects of the spectrum which are
of particular interest. Before discussing considerations for detecting
the back-action noise, we compare the spectrum with that which would
have been obtained using the SBMME, and for that which would result
from the use of homodyne detection. We then write the spectrum at
resonance as a function of the laser power, and plot this for current
experimental parameters. So far we have been considering the noise
power spectrum, and have made no particular reference to the limit this
implies for a measurement of the position of the mirror. In
section~\ref{partc} we show how the spectrum tells us the limit to the
accuracy of position measurement in the presence of the noise sources.

\subsection{Comparison with the Standard Treatment of Brownian Motion}
To obtain the spectrum we have used the corrected Brownian motion
master equation~\cite{Diosi}. This is essential because the spectrum
which results from the standard Brownian motion master equation
contains a term which is assymetric in $\omega$, and therefore clearly
incorrect. In particular, to obtain the spectrum given by the SBMME
from that given by the CBMME, the term proportional to
$T_{\mbox{s}}^{-1}$ must be replaced by
\begin{equation}
   2\omega\gamma\Gamma\chi^2\alpha^2\nu \left[ \frac{\left( \left(
\frac{\gamma+\mu}{2}\right)^2 + \omega^2 \right)}  {|D(\omega)|^2}
\right].
\end{equation}
That the spectrum must be symmetric in $\omega$ follows readily from
the stationarity of the output field, and the fact that the output
field commutes with itself at different times. In particular, the
stationarity of the output field means that the correlation function
of the signal only depends on the time difference, so that
\begin{equation}
  \langle R(t)R(t+\tau)\rangle = G(\tau) .
\end{equation}
As the output field commutes with itself at different times, $R$
commutes with
itself at different times, and we have
\begin{equation}
  G(-\tau) = \langle R(t)R(t-\tau)\rangle = \langle
R(t-\tau)R(t)\rangle
           = G(\tau) .
\end{equation}
The correlation function is therefore symmetric in $\tau$. As the
spectrum is the Fourier transform of the correlation function, it
follows from the properties of the Fourier transform that the
spectrum is symmetric in $\omega$.

It was shown in Ref.~\cite{Haake} that for realistic systems at 
high temperatures the SBMME has a stationary density matrix which is 
positive. The non-Lindblad nature of the master equation appears only 
to cause problems at short times. In our problem we are calculating 
spectra at steady-state so it might seem surprising that the 
non-Lindbad nature does cause problems for us. On reflection, however,
this is not surprising. The spectra we calculate are for continuously 
measured quantities. Making such measurements continuously reprepares 
the system in a {\em conditioned} state which is different from the 
stationary state. Thus if one is observing the system then it is never 
really at steady state and the ``initial slip'' problem of 
Ref.~\cite{Haake} never goes away.

Diosi's corrected Brownian motion master equation removes the term
asymmetric in $\omega$ by adding a noise source to the position (see
Eq.(\ref{qle3})) which is correlated with the noise source for the
momentum. In doing so it produces an additional term in the spectrum
proportional to $1/T$, an effect which, it should be noted, is
independent of the phase detection scheme. For temperatures (and
frequencies) for which this new term is much smaller than the standard
term, which is proportional to $T$, this new term can be neglected.
However, the question of observing this term experimentally is a very
interesting one, because it would allow the CBMME to be tested.
Comparing the new term with the term proportional to $T$ we find that
the new term begins to dominate when
\begin{equation}
T < \left( \frac{\hbar}{12 k_{\mbox{\scriptsize B}}} \right)
\sqrt{\Gamma^2 + \omega^2} .
\end{equation}
For temperatures of the order of a few Kelvin, the additional term
therefore becomes apparent in the spectrum at frequencies of a few
Gigahertz. Note that for such high frequencies phase modulation may no
longer be practical however, owing to the fact that $\Delta$ must be
much larger than the frequency range of the signal. In that case
the use of alternative phase detection schemes would be required

\subsection{Comparison with Homodyne Detection}
Let us now briefly compare the spectrum derived above for phase
modulation detection to that which would be obtained with homodyne
detection. Firstly, if homodyne detection had been used, the
overall scaling of the spectrum would be different,
as it would be proportional to the strength of the local oscillator.
Thus the factor of $1/(\varepsilon\beta)^2$ would be
replaced by $1/(\tilde{\beta}\kappa)^2$, in which $\tilde{\beta}$
and $\kappa$ are as defined in Eq.~(\ref{above}). This
overall factor aside, two terms in the spectrum would change. The
shot noise component would be reduced to unity, and the classical
phase noise contribution would become
\begin{equation}
  4\tilde{G}_y(\omega) \left[ \frac{\left( \frac{\gamma -\mu}{2}
\right)^2 + \omega^2}{ \left( \frac{\gamma+\mu}{2} \right)^2 +
\omega^2} \right] .
\end{equation}

\subsection{The Error in a Measurement of Position}
\label{partc}
So far we have been considering the noise spectrum of the phase
quadrature, as this is what is actually measured. In this
section we show how the error in a measurement of the position of
the mirror may be obtained in a simple manner from the spectrum,
Eq.(\ref{eqpsd}), and give an example by calculating it explicitly
for some of the terms. As explained above, the reason for performing the
phase measurement is that it constitutes essentially a measurement of
the position of the mirror.

We can choose to measure the amplitude of position oscillations at
any frequency, but for the purposes of discussion, a measurement of
a constant displacement is the simplest. First we must see how the
position of the mirror appears in the signal, which is the phase
quadrature measurement (that is, convert from the units of the
signal into units of the position fluctuations). This is easily
done by calculating the contribution to the spectrum of the position
fluctuations due to one of the noise sources (for the sake of
definiteness we will take the thermal noise), and comparing this to the
equivalent term in the spectrum of the signal. This gives us the
correct scaling. Performing this calculation, we find that the
spectrum of position fluctuations of the mirror due to thermal noise
is given by the thermal term in the spectrum (Eq.(\ref{eqpsd})),
multiplied by the factor
\begin{equation}
 \frac{\hbar}{2m(\varepsilon\beta)^2\gamma\nu\chi^2\alpha^2}
\left(\left(
 \frac{\gamma+\mu}{2} \right)^2 + \omega^2 \right) .
\label{poss}
\end{equation}
From this we see that the scaling factor is frequency dependent. This
means,that the spectrum of the position fluctuations is somewhat
different from the spectrum of the resulting phase quadrature
fluctuations. For the measurement of the phase to correspond to a true
measurement of the position the two spectra should be the same. This is
true to a good approximation when $\gamma$ is much larger than the
range of $\omega$ over which the spectrum of position fluctuations is
non-zero, and this is why the scheme can be said to constitute a
measurement of position when $\gamma\gg\nu,\Gamma$.

In performing a measurement of a constant displacement of the mirror
(achieved by some constant external force), the signal (after scaling
appropriately so that it corresponds to position rather than
photocurrent)
is integrated
over a time $\tau_{\mbox{\scriptsize m}}$. The best estimate of the
displacement
is this integrated signal divided by the measurement time. The error,
$\Delta x$, in the case that the measurement time is much greater
than the
correlation time of the noise, is given by
\begin{equation}
  \Delta x^2(0) = \int_{-\infty}^{\infty} \!\!\!\! \langle
R_x(0)R_x(\tau)\rangle d\tau
               /\tau_{\mbox{\scriptsize m}} =
S_x(0)/\tau_{\mbox{\scriptsize m}}.
\end{equation}
In this equation $R_x$ and $S_x$ are the appropriately scaled signal and
spectrum. To calculate the error in the measurement of a constant
displacement, all we have to do, therefore, is to scale the spectrum
using the expression Eq.(\ref{poss}), evaluate this at zero frequency,
and divide by the measurement time. In general, the spectrum evaluated
at a given frequency, once divided by the measurement time, gives the
error in a measurement of the amplitude of oscillations at that
frequency. We calculate now the contribution to the error in a
measurement at zero frequency and at the mirror resonance frequency,
from the shot noise, thermal, and quantum back-action noise. In the
following we write the expressions in terms of the parameters usually
used by experimentalists: the laser power, $P$, cavity finesse ${\cal
F}$, and the quality factor for the mirror oscillator, $Q=\nu/\Gamma$.
We chose the cavity to be impedance matched, since this is usually the
case in practice. This means that the decay rate due to the input
coupler, $\gamma$, is chosen equal to the internal cavity decay rate,
$\mu$. The total decay rate of the cavity is therefore $2\gamma$, so
that the finesse is given by ${\cal F}= \pi c/(2L\gamma)$. We also
assume that $\gamma \gg \nu$, which is certainly true in current
experiments. Performing the calculation we find that the contribution
due to the shot noise is the same at all frequencies, and is given by
\begin{equation}
     \Delta x^2_{\mbox{\scriptsize SN}} = \frac{3\pi^2}{32} \left(
\frac{\hbar c^2}{\omega_0} \right) \frac{1}{{\cal F}^2 P
\tau_{\mbox{\scriptsize
m}}} .
\end{equation}
The contribution from the quantum back action for a measurement of a
constant displacement is
\begin{equation}
      \Delta x^2_{\mbox{\scriptsize BA}}(0) = \frac{4}{\pi^2} \left(
\frac{\hbar\omega_0}{c^2} \right) \left( \frac{1}{m^2\nu^4} \right)
\frac{{\cal F}^2 P}{\tau_{\mbox{\scriptsize m}}},
\end{equation}
and for a measurement at the resonance frequency $\nu$ it is
$\Delta x^2_{\mbox{\scriptsize BA}}(\nu) = Q^2
\Delta x^2_{\mbox{\scriptsize BA}}(0)$. Note that since $\mu=\gamma$,
the contribution from the internal cavity losses is also given by this
expression. In a sense, the internal cavity loss noise can also be
regarded as a back-action term, although the back action is from a
measurement process due to the interaction with an environment that is
not being observed. The total error which can be said to arise from the
random `photon impacts' on the mirror (in the absence of classical laser
noise) is the sum of the back-action and internal loss noise, and is
therefore given by
\begin{equation}
      \Delta x^2_{\mbox{\scriptsize PN}}(0) = \frac{8}{\pi^2} \left(
\frac{\hbar\omega_0}{c^2} \right) \left( \frac{1}{m^2\nu^4} \right)
\frac{{\cal F}^2 P}{\tau_{\mbox{\scriptsize m}}}.
\end{equation}
The contribution from the thermal noise is
\begin{equation}
      \Delta x^2_{\mbox{\scriptsize Th}}(0) =  \left(
\frac{2 k_{\mbox{\scriptsize B}}T}{m\nu^3 Q \tau_{\mbox{\scriptsize
m}}} \right) + \left( \frac{\hbar^2}{6m\nu k_{\mbox{\scriptsize
B}}TQ^3 \tau_{\mbox{\scriptsize m}}} \right),
\end{equation}
for a constant displacement, and is
\begin{equation}
      \Delta x^2_{\mbox{\scriptsize Th}}(\nu) =  \left(
\frac{2 k_{\mbox{\scriptsize B}}QT}{m\nu^3 \tau_{\mbox{\scriptsize
m}}} \right) + \left( \frac{\hbar^2 Q}{6m\nu k_{\mbox{\scriptsize
B}}T \tau_{\mbox{\scriptsize m}}} \right),
\end{equation}
for an oscillation at the mirror frequency. In obtaining the second
term in this last expression we have also used 

\begin{figure}
\leavevmode
\epsfxsize=8cm
\epsfbox{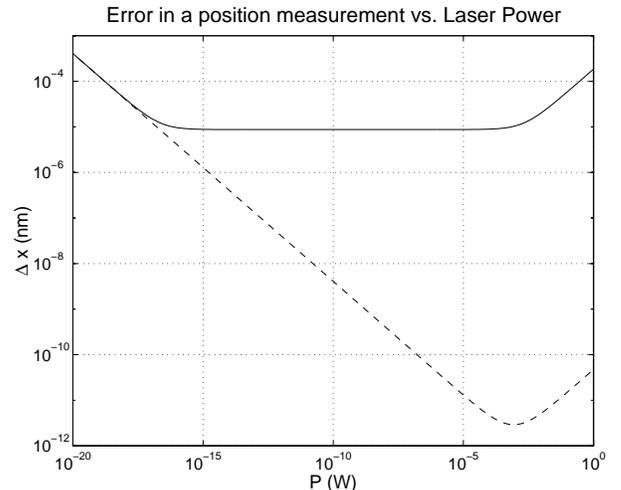}
\caption{\narrowtext The error in a measurement of the position of
the mirror, for a measurement time of $\tau_{\scriptsize m}=300$s.
The dashed curve corresponds to a measurement of a constant
displacement, and the dot-dash curve to a measurement at the
mirror resonance frequency. The combined contribution of the 
quantum back-action and internal cavity loss noise to both curves 
is the sloping section to the right. The quantum back-action noise may
be observed at reasonable laser power levels. For parameters see text.}
\label{theoryfig2}
\end{figure}

$\!\!\!\!\!\nu \gg \Gamma$. The
contribution from the other noise sources may also be readily
evaluated from the terms in the spectrum Eq.(\ref{eqpsd}).

Let us examine the total error in a position measurement resulting
from these four contributions (shot-noise, back-action, internal
losses, and thermal noise) for state-of-the-art experimental
parameters. Reasonable values for such parameters are as
follows~\cite{Tittonen98}. The laser frequency is
$\omega_0=2\pi\times 2.82\times 10^{14}\;\mbox{rad s}^{-1}$ (assuming
a Nd:YAG laser with a wavelength of 1064 nm), the cavity length is
$L=1\;\mbox{cm}$, the mass of the oscillating mirror is
$m=10^{-5}\;\mbox{kg}$, and the resonant frequency of the mirror is
$\nu=2\pi\times 2\times 10^4\;\mbox{rad s}^{-1}$. The quality factor
of the mirror is four million, which gives $\Gamma \approx 3\times
10^{-2}\;\mbox{s}^{-1}$. With these parameters for the cavity we have
$\chi=2.29\times 10^{-4}\;\mbox{s}^{-1}$. The cavity damping rate
through the front mirror is $\gamma = 4.7 \times 10^5 \;\mbox{s}^{-1}$,
and we assume impedance matching so that $\mu=\gamma$. The
cavity may be cooled to a temperature of $T=4.2\;\mbox{K}$, so that
$T_{\mbox{s}}=k_{\mbox{\scriptsize B}}T/(\hbar\nu)=4.37\times 10^6$,
which is certainly in the high temperature regime ($T_{\mbox{s}}\gg
1$). The Diosi term (of order $T_{\mbox{s}}^{-2}$ at resonance) is thus
totaly negligible.

In figure~\ref{theoryfig2} we plot the position measurement error as
a function of the laser power, both for the measurement of a constant
displacement, and for a displacement at the mirror resonance
frequency. The expressions for the measurement error derived above
are valid in the limit where the measurement time is much greater than
the correlation time of the noise. As the cavity-mirror system is
driven by white noise, this correlation time is given approximately
by the longest decay time of the system. In our case this is the
decay time of the moving mirror, given by $1/\Gamma \approx
30\;\mbox{s}$. In view of this we have chosen a measurement time of
300 seconds (5 minutes) for the plot in figure~\ref{theoryfig2}.

The uncertainty due to the shot noise falls off with laser power,
while that due to thermal noise is independent of laser power, and
that due to the quantum back-action increases with laser power.
These results are already well known. The thermal and back-action
contributions are much greater at the resonance frequency of the
mirror, due to the high mechanical $Q$ factor. The optimal regime for
detecting the quantum back-action noise is at resonance, as the
absolute magnitude of this noise is largest in this case. Reasonable
experimental values for laser power lie between the solid lines, where
the increase in noise due to the back-action is visible. However, our
analysis of the spectrum shows us that the full situation is more
complicated. We have shown that the noise due to internal cavity losses
and the classical laser amplitude noise have the same dependence on
frequency as the quantum back-action. In order to reach the back-action
dominated regime, the laser amplitude noise must be at the shot noise
level, and the frequency noise must be extremely low.

\section{Conclusion}\label{sec6}

We have examined the optomechanical system consisting of a Fabry-Perot
cavity containing a moving mirror to see how the quantum mechanical
back-action appears among the various sources of classical noise. We
have shown a number of things regarding this question. First of all,
the relationship of the shot noise to the noise resulting from the
oscillating mirror, and hence the limit on a position measurement
due to the shot noise, is dependent on the phase measurement scheme.
In particular, the result for phase modulation detection, which is
commonly used in experiments of this kind, is not the same as that for
homodyne detection. We have found that while the signature of the
classical phase noise is quite different for that of the quantum-back
action, the noise due to intracavity losses and classical amplitude
noise has a very similar signature to the back-action. As far
as the parameters of the cavity and oscillating mirror are concerned,
realisable experiments are beginning to fall in the region where the
quantum back-action may be observed.

In our treatment of the system we have shown that the standard
quantum Brownian motion master equation produces a clearly spurious
term in the steady state noise spectrum for the phase quadrature
measurement. We have shown that the corrected Brownian motion master
equation, derived by Diosi, corrects this error. However, it also
produces a new term in the spectrum which is small for present
experimental systems. Testing for the existence of this term poses an
experimental challenge that might be met using miniature, high
frequency oscillators and ultra-low temperatures.


\section*{Acknowledgements}

Discussions on the subject of quantum limited measurements
with J\"urgen Mlynek, Gerd Breitenbach, Thomas M\"uller, and Thomas
Kalkbrenner at the University of Konstanz are gratefully acknowledged.
I.T. wishes to thank the Alexander von Humboldt Foundation, and J.
Mlynek for his hospitality. K.J. wishes to acknowledge the support of
the British Council and the New Zealand Vice Chancellors Committee.
H.W. was supported by the Australian Research Council and the
University of Queensland.

\end{multicols}\widetext

\begin{multicols}{2}

\end{multicols}

\end{document}